\documentclass[12pt]{article}
\usepackage{graphicx}
\usepackage{lineno}

\newcommand\pubnumber{}
\newcommand\pubdate{\today}

\def\napoli{University College Dublin, Dublin 4, Ireland}

\def\Title#1{\begin{center} {\Large #1 } \end{center}}
\def\Author#1{\begin{center}{ \sc #1} \end{center}}
\def\Address#1{\begin{center}{ \it #1} \end{center}}

\newcommand\pubblock{\rightline{\begin{tabular}{l} \pubnumber\\
         \pubdate  \end{tabular}}}
\newenvironment{Abstract}{\begin{quotation}  }{\end{quotation}}
\newenvironment{Presented}{\begin{quotation} \begin{center} 
             PRESENTED AT\end{center}\bigskip 
      \begin{center}\begin{large}}{\end{large}\end{center} \end{quotation}}

\def\beq{\begin{equation}}
\def\eeq#1{\label{#1}\end{equation}}
\def\eeqn{\end{equation}}

\def\beqa{\begin{eqnarray}}
\def\eeqa#1{\label{#1}\end{eqnarray}}
\def\eeqan{\end{eqnarray}}

\let\bar=\overbar

\def\Dslash{\not{\hbox{\kern-4pt $D$}}}
\def\dslash{\not{\hbox{\kern-2pt $\del$}}}

\def\msb{{\bar{\ssstyle M \kern -1pt S}}}

\begin{document}
\begin{titlepage}
\pubblock

\vfill
\Title{Central Exclusive Production at LHCb}
\vfill
\Author{Ronan McNulty (on behalf of the LHCb collaboraiton)}
\Address{\napoli}
\vfill
\begin{Abstract}
The installation of a new sub-detector has improved the ability of LHCb to measure central
exclusive production in Run 2 at the LHC.  A measurement of central exclusive $J/\psi$ production
is presented and improvements for the analysis of hadronic final states are discussed. 
\end{Abstract}
\vfill
\begin{Presented}
Presented at EDS Blois 2017, Prague, \\ Czech Republic, June 26-30, 2017
\end{Presented}
\vfill
\end{titlepage}
\def\thefootnote{\fnsymbol{footnote}}
\setcounter{footnote}{0}

\section{Introduction}

Central Exclusive Production (CEP) in hadron collisions occurs when colourless propagators are exchanged between
the hadrons creating a central system, with the colliding hadrons remaining intact after the interaction.  This leads to the unique experimental signature of a small number of particles in the central system with rapidity gaps extending
to the projectiles.  LHCb has performed measurements of the exclusive differential cross-sections as 
a function of rapidity of $J/\psi$, $\psi(2S)$
and $\Upsilon$ mesons~\cite{jpsi7,ups78} in $pp$ collisions at $\sqrt{s}$ of 7 and 8 TeV.  
These vector mesons are created through photon-Pomeron fusion and the measurements probe the gluon PDF down to fractional parton momenta of $2\times10^{-6}$~\cite{pdf,pdf2}, 
being potentially sensitive to saturation effects~\cite{sat} or the existence of odderons~\cite{odd}, colourless three-gluon states.
LHCb has also measured double charmonia production~\cite{djpsi} in Pomeron-Pomeron fusion.
This shows the complementarity of measurements through CEP and conventional QCD production~\cite{djpsiqcd},
and demonstrates that CEP has potential to search for tetraquarks and exotic
or hybrid mesons.  In particular, double Pomeron exchange, being rich in gluons, is a good laboratory in which to search for glueball states.

In these proceedings, I discuss improvements made by LHCb in the collection and understanding of CEP events
during Run 2, present a recent measurement of charmonia production in $pp$ collisions at $\sqrt{s}=13$ TeV, and
outline the future potential of CEP measurements.

\section{Central Exclusive Production at Run 2}

Experience of CEP analyses in Run 1 at the LHC showed it was relatively easy to select events in the LHCb detector
that  had no other activity apart from the particles making up the central system.  However, it was more difficult to determine whether these were truly exclusive, or whether the protons dissociated with the remnants travelling outside the acceptance of the LHCb detector.  Discrimination was provided by considering the transverse momentum, $p_T$, of the central system.  
The distribution of events as a function of the Mandelstam variable, $t\approx p_T^2$ was assumed to follow a distribution $\exp(bt)$ where the slope $b$  is greater for signal than for background: thus a fit to data can extract
the relative proportion of exclusive events.

During Run 1, the LHCb detector was fully instrumented in the pseudorapidity region $2<\eta<5$, and could register the presence of charged tracks in the region $-3.5<\eta<-1.5$.  In order to extend the region where particle activity could be recorded, a new sub-detector, HeRSCheL~\cite{herschel}, consisting of five planes of scintillators was installed on both sides of LHCb in the LHC tunnel, during the LHC shutdown following Run 1.  It adds a veto region of $5<|\eta|<10$ to the acceptance of LHCb for Run 2.

\begin{figure}[htb]
 \centering
 \includegraphics[height=2.5in]{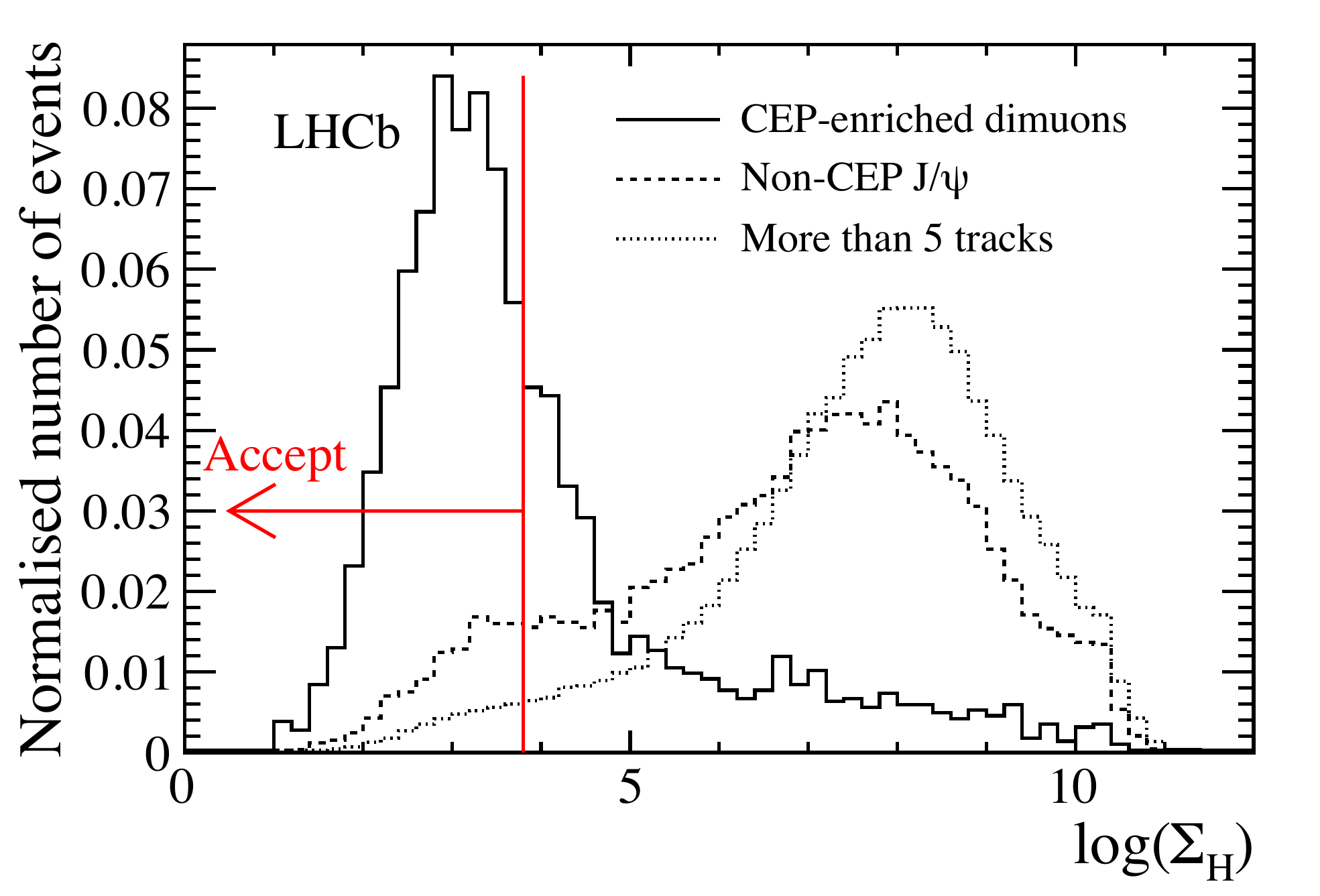}
 \caption{Summed energy in the HeRSCheL detector for three classes of dimuon events: 
candidates for CEP from the QED continuum with $p_T^2<0.01$ GeV$^2$; 
$J/\psi$ non-exclusive candidates with $p_T^2>1$ GeV$^2$; and low multiplicity events with five or more
reconstructed tracks.  The vertical line indicates the region used in the measurement of the $J/\psi$ 
and $\psi(2S)$ cross-sections
at $\sqrt{s}=13$ TeV.}
 \label{fig:sigmah}
\end{figure}

Each station of HeRSCheL consists of four scintillators, 30~cm square, arranged symmetrically around the beam-pipe and read out by photomultiplier tubes.  The digitised signals undergo a callibration procedure and the summed signal, $\Sigma_H$, is used as a discriminant.  The response of HeRSCheL to three classes of events is shown in 
Fig.~\ref{fig:sigmah}.  The first class is enriched in CEP events, which have been selected from dimuon candidates having $p_T^2<0.01$ GeV$^2$ and masses greater than 1.5 GeV, but outside the charmonium or bottomonium regions.
The second class consists of $J/\psi$ candidates having $p_T^2>1$ GeV$^2$ and are thus likely to be non-exclusive.
The third class has events triggered by a low-multiplicity muon trigger but with 5 or more charged tracks.  The activity
observed in HeRSCheL for exclusive events is lower than, and shows a clear distinction from, 
the activity in inclusive events.  Therefore, requiring low activity in HeRSCheL
increases the fraction of exclusive events.

Events with low multiplicity were selected with a two-stage trigger.  The first stage, L0, was
a hardware trigger and required low multiplicity in the scintillating pad detector in
conjunction with deposits in the muon chambers, electromagnetic or hadron calorimeters.
The second stage was implemented in software.
Improvements to the trigger in Run 2 included a lowering of the thresholds so that mesons produced
with $p_T=0$ GeV could be selected by the muon, electromagnetic and hadronic triggers if the mass of the meson was above 700, 300 and 400 MeV, respectively.

\section{Exclusive production of $J/\psi$ and $\psi(2S)$ mesons}

\begin{figure}[htb]
 \centering
 \includegraphics[height=3in]{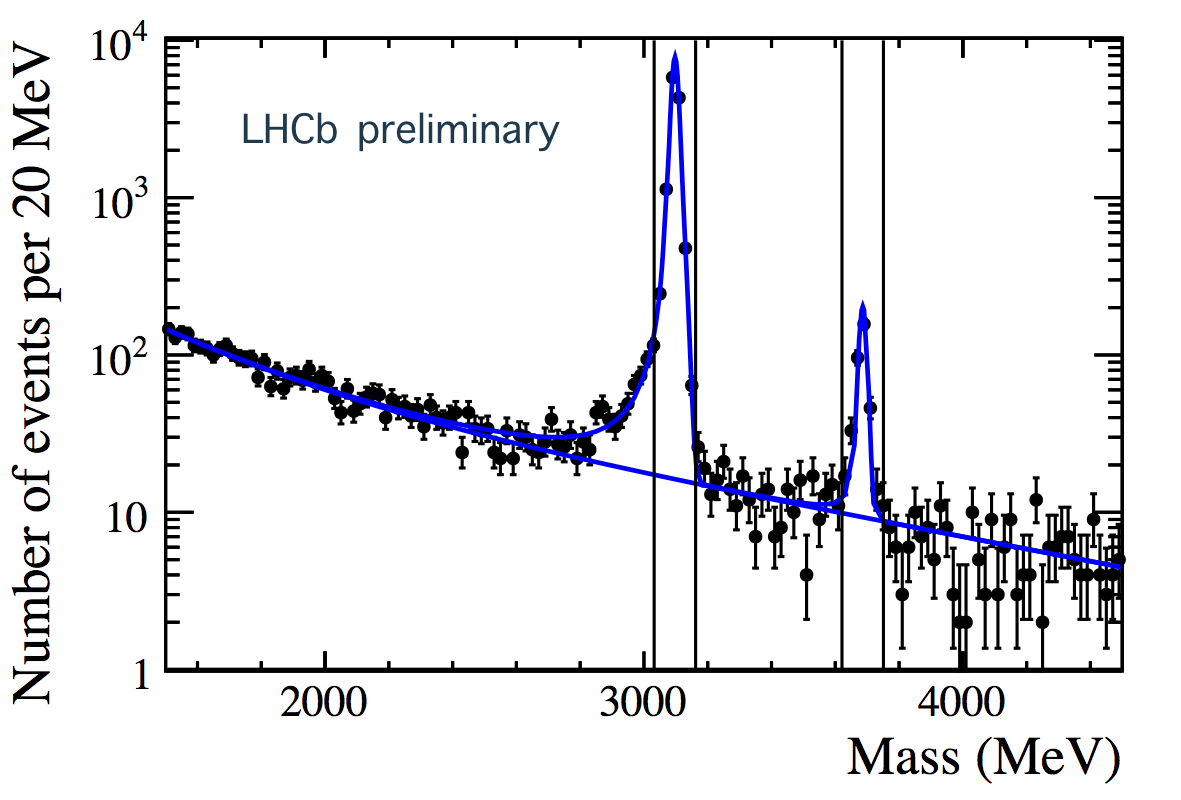}
 \caption{Invariant mass of exclusive dimuon candidates.  The lines are fits to the data.  The windows
indicate the regions chosen for the $J/\psi$ and $\psi(2S)$ analyses.}
 \label{fig:mumass}
\end{figure}

\begin{figure}[htb]
 \centering
 \includegraphics[height=2in]{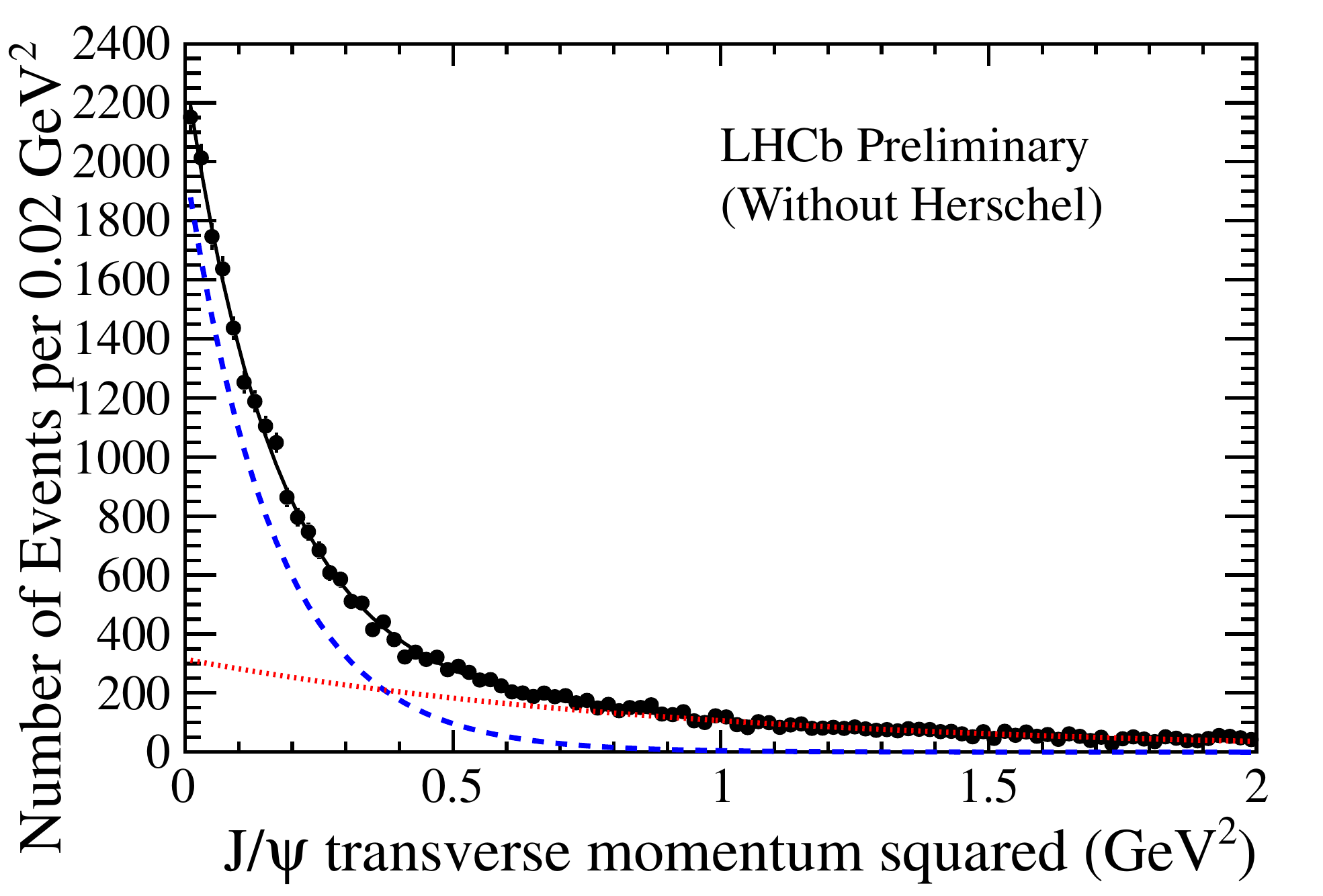}
 \includegraphics[height=2in]{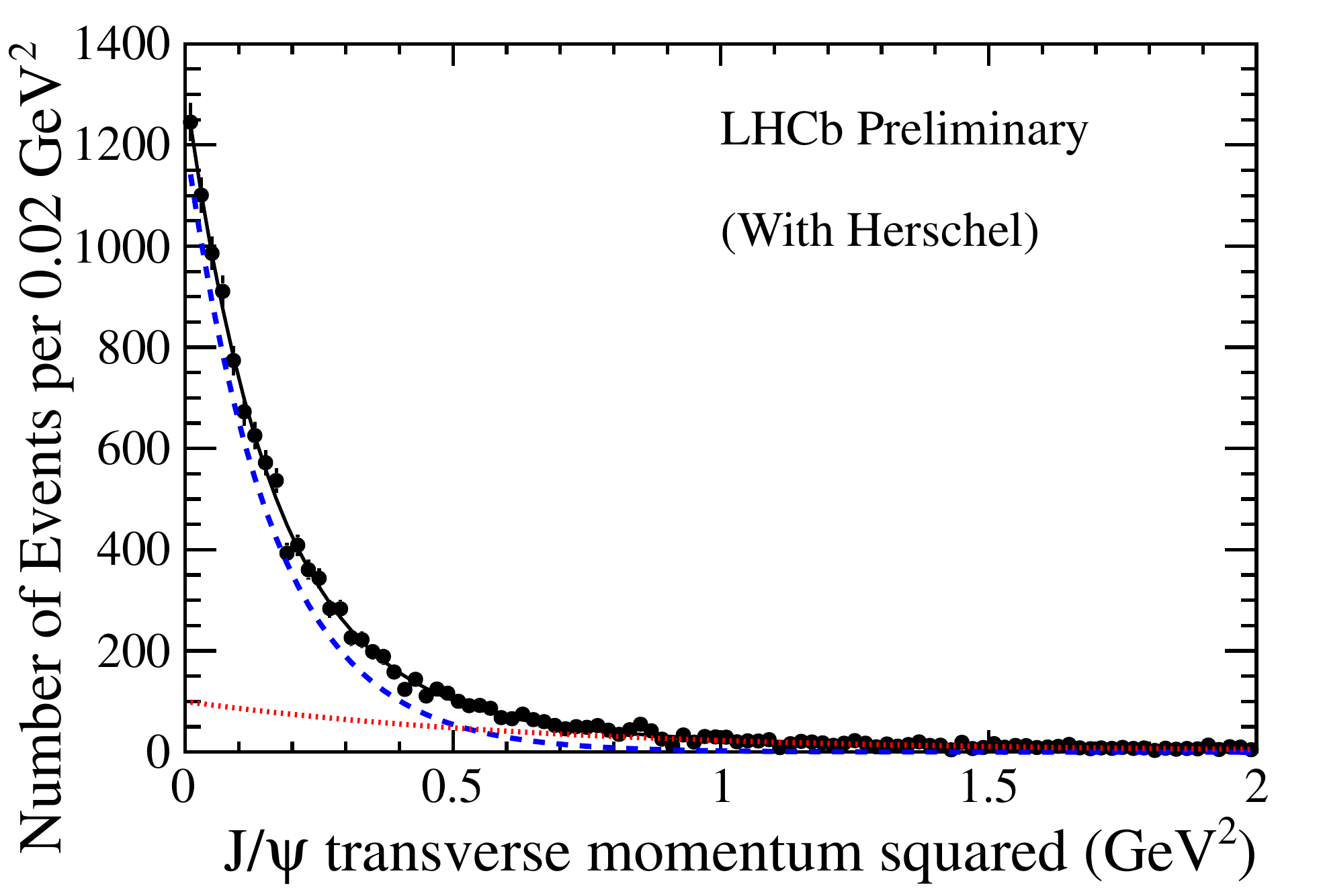}
 \caption{Transverse momentum squared of $J/\psi$ candidates without (left) and with (right) the requirement
that there be little activity in HeRSCheL.  The solid line represents the total fit.
The dashed line is the estimated exclusive component while the  
dotted line is the proton dissociation.}
 \label{fig:pt2}
\end{figure}

The selection of exclusive charmonia produced at $\sqrt{s}=13$ TeV~\cite{jpsi13} and decaying to two muons 
was similar to the analysis published at 7 TeV~\cite{jpsi7}, but with the addition of requiring little activity in the
newly installed HeRSCheL sub-detector, as indicated by the vertical line in Fig.~\ref{fig:sigmah}.  Precisely two muons were required in the event with no additional charged or neutral particles reconstructed.  The invariant mass 
of the two muons is shown in Fig.~\ref{fig:mumass} and candidates for $J/\psi$ and $\psi(2S)$ were required to be within
65 MeV of their known masses.  Contamination in the $J/\psi$ sample due to feed-down from $\chi_c\rightarrow J\psi\gamma$ where the photon escapes detection was estimated by scaling, using simulation, the number of 
reconstructed $\chi_c$ events where the photon was observed.
A fit to the $p_T^2$ distribution is sensitive to the relative amount of exclusive and
non-exclusive events and is shown in Fig.~\ref{fig:pt2} before and after the HeRSCheL requirement.
The new sub-detector roughly halves both the feed-down and proton-dissociation components, reducing the
systematic uncertainties associated with knowledge of the backgrounds.

\begin{figure}[htb]
 \centering
 \includegraphics[height=2.in]{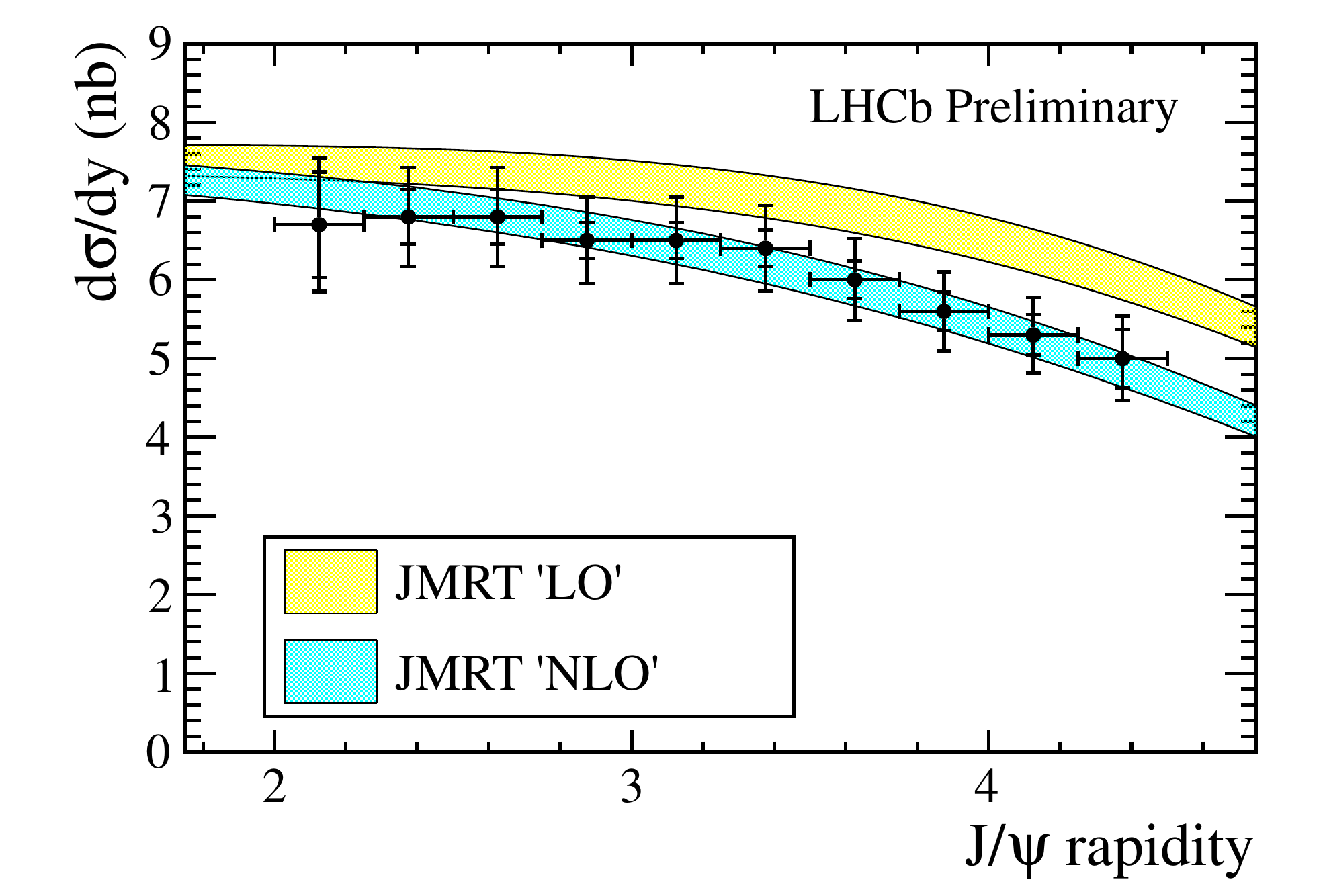}
 \includegraphics[height=2.in]{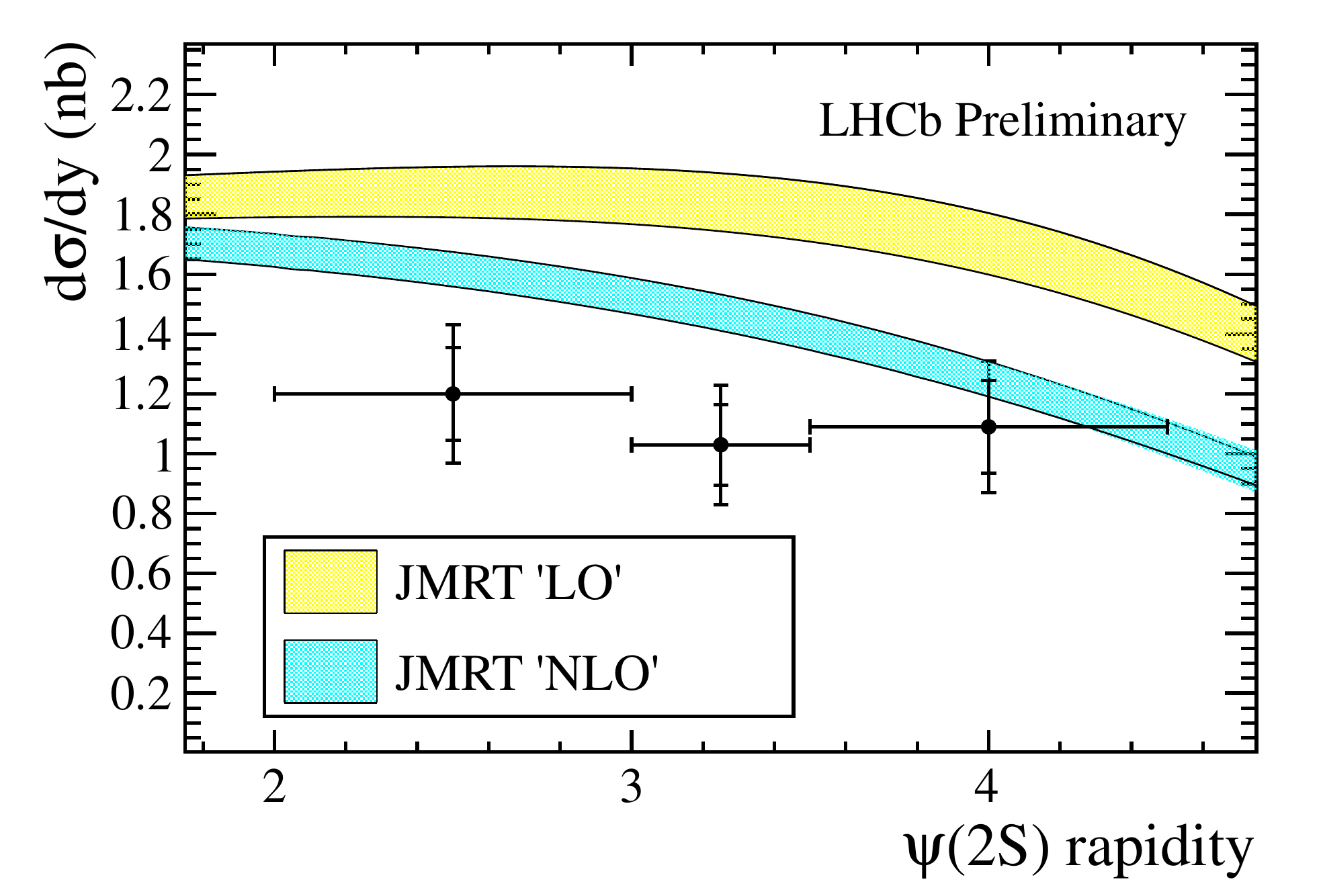}
 \caption{Differential cross-section for $J/\psi$ (left) and $\psi(2S)$ (right) production at $\sqrt{s}=13$ TeV compared
to theoretical predictions.}
 \label{fig:cs}
\end{figure}

The efficiency for the selection was calculated using data.  The efficiency for triggering, tracking and the muon selection
used a tag-and-probe technique where one of the two muons was `tagged' and the efficiency for the other was measured.  The efficiency for the requirement on activity in HeRSCheL was found from the QED-produced dimuon
sample shown in Fig.~\ref{fig:sigmah}.  The purity was obtained from the fit shown in Fig.~\ref{fig:pt2}.  
The cross-section was obtained from the number of events produced in the estimated integrated luminosity 
of $204\pm8$ pb$^{-1}$.  The differential cross-section as a function of rapidity for both 
$J/\psi$ and $\psi(2S)$ are shown in Fig.~\ref{fig:cs} compared to approximate NLO theoretical calculations~\cite{jpsith}.

\section{Central exclusive production of pion pairs}

The new triggers installed for Run 2 have allowed LHCb to collect a large sample of pion pairs with no other detector
activity in proton-proton and proton-lead collisions.  The dominant production mechanism is double
pomeron exchange in the former while in the latter it is photoproduction with the pions produced from the decay of the
$\rho$ meson.  Photoproduction of the $\rho$ on protons was measured at HERA up to photon-proton
centre-of-mass energies of about 200 GeV~\cite{herarho}.  However,
LHCb can access values up to 1 TeV, corresponding
to a parton fractional momentum, $x=1\times 10^{-6}$.  
At such small $x$ values, saturation effects might contribute, and these
would lead to a change in the cross-section rise measured at HERA.

\begin{figure}[htb]
 \centering
 \includegraphics[height=1.8in]{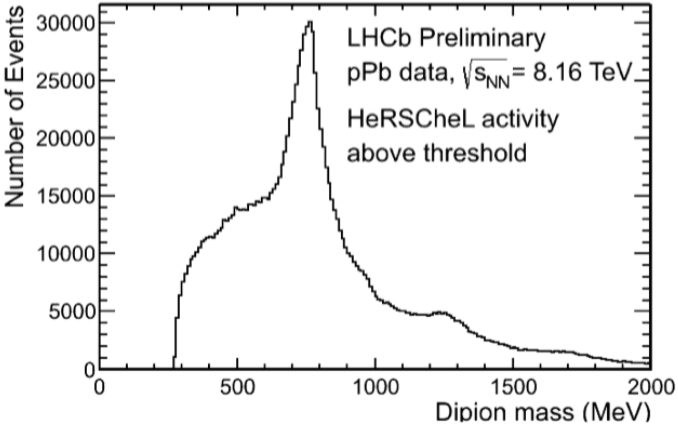}
 \includegraphics[height=1.8in]{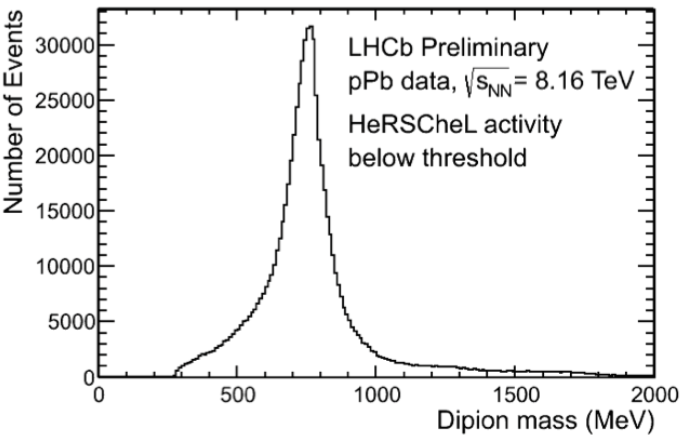}
 \caption{Candidates for exclusive dipion production in proton-lead collisions requiring activity in HeRSCheL
above (left) and below (right) a threshold.}
 \label{fig:rho}
\end{figure}

A sample of exclusive dipions was selected by requiring precisely two charged tracks coming from the interaction region, consistent with the pion
hypothesis (mainly coming from information in the RICH sub-detector), 
and with no other neutral or charged deposits in the LHCb detector.
These events appear exclusive in LHCb but may not be truly exclusive if there exist
other particles outside the LHCb acceptance.  
However, the HeRSCheL sub-detector allows activity in a much wider pseudorapidity region to be vetoed. 
Using the response of the detector in QED-produced dimuon events, a threshold was defined that was 
compatible with no
activity in HeRSCheL.  The invariant mass distribution for dipion events below and above this threshold is shown
in Fig.~\ref{fig:rho}.    The rho resonance is clear in both distributions, but the activity outside the resonance region
is strongly suppressed when no activity is present in HeRSCheL.  This suggests that the events without HeRSCheL
activity are enriched in central exclusive production.

\section{Conclusions}
The installation of the HeRSCheL subdetector for Run 2 has increased the sensitivity of LHCb to central exclusive production.  The cross-section for exclusive $J/\psi$ production has been measured at $\sqrt{s}=13$ TeV.  
Using HeRSCheL together 
with improvements to the trigger, a large sample of central exclusive produced hadronic final states now awaits analysis.
In addition to the potential to measure $\rho$ production, pion, kaon and proton
pairs can be measured in both proton-proton and proton-ion collisions thus 
allowing an investigation of a wide variety of light meson and charm spectroscopy in both photoproduction
and double pomeron exchange processes.  Such spectroscopy is important in searches for glueballs and
exotic hadrons as well as for new QCD phenomena such as odderons and saturation.




\end{document}